\documentstyle[epsf,aps,preprint]{revtex}
\newcommand{\scalegraphlabel}[4]{
\begin{figure}
\epsfxsize=#2
\centerline{\epsfbox{#1}}
\caption{\footnotesize #3}
\label{#4}
\end{figure}
}

\begin{document}
\draft
\title{Friction Drag on a Particle Moving in a Nematic Liquid Crystal}
\author{R. W. Ruhwandl and E. M. Terentjev}
\address{Cavendish Laboratory, University of Cambridge, Madingley
Road, Cambridge CB3 0HE, U.K.}
\date{\today}

\maketitle

\begin{abstract}
  The flow of a liquid crystal around a particle does not only depend
  on its shape and the viscosity coefficients but also on the
  direction of the molecules. We studied the resulting drag force on a
  sphere moving in a nematic liquid crystal (MBBA) in a low Reynold's
  number approach for a fixed director field (low Ericksen number
  regime) using the computational artificial compressibility method.
  Taking the necessary disclination loop around the sphere into
  account, the value of the drag force anisotropy
  ($F_\perp/F_\parallel=1.50$) for an exactly computed field is in
  good agreement with experiments ($\sim1.5$) done by conductivity
  diffusion measurements. We also present data for weak anchoring of
  the molecules on the particle surface and of trial fields, which
  show to be sufficiently good for most applications.  Furthermore, the
  behaviour of the friction close to the transition point
  nematic$\leftrightarrow$isotropic and for a rod-like and a disc-like
  liquid crystal will be given.

\end{abstract}

\vspace{0.4in}

\pacs{PACS: 61.30.Jf; 61.30.Cz; 83.85.Pt}

\section{Introduction}

Most of the applications of liquid crystals are connected to their
flow properties, either while processed or in the application
itself. The reorientation of the director field for example, which is
used in electro-optical devices, is linked to internal flow. The
fastest response times needed for a further development are limited by
the friction effects. But although the basics of the hydrodynamics of
liquid crystals were laid about thirty years ago, most of the problems
connected with flow are still unsolved. This is mainly due to the
anisotropy of the system and the non-trivial connection of the
direction of the molecules and the velocity.

A deeper insight in the hydrodynamics of liquid crystals and the
connection between macroscopic and microscopic properties would allow
to predict the behaviour of particular materials and therefore to
design special liquid crystals to obtain certain characteristics
required.

Precision experiments are often difficult to perform, since many of
the standard techniques do not work for these materials. It would be
useful to have more independent methods of measuring the viscosity,
than only a traditional shear flow. A further technique, the falling
ball experiment, was solved for an isotropic liquid by Stokes. It
consists of a ball falling down in a cylinder driven by the
gravitational force and measuring its equilibrium velocity. The
viscosity $\eta$ can then be determined by the well known Stokes'
formula $F_D=-6\pi r\eta v$, which gives the relation between the
friction drag $F_D$, the radius $r$ of the ball and its velocity $v$.

For liquid crystals this problem gets another dimension since the drag
force also depends on the geometry of the system. It is obvious that
the drag force on the sphere is different for the two particular cases
of flow and director parallel and flow and director perpendicular to
each other. Using a liquid crystal of rod-like molecules it becomes
clear that it is easier to move the particle parallel to the general
director field, i.e.  along the long axis of the molecules, than to
move it perpendicular to the director, i.e.  perpendicular to the long
axis.  In the general case (arbitrary angle between flow and director)
this results in the fact, that the drag force is no longer parallel to
the line of motion. There is a further component perpendicular to it,
the so called lift force, which moves the particle sideways (see
fig.\ref{dir_field_force}). It is worth mentioning that this force
does not contribute to the dissipative losses in the system, an effect
well known from other areas of physics like electro-dynamics (a
charged particle in a magnetic field is forced to change its direction
without loosing or gaining any kinetic energy).

A further problem is the influence of the director field ${\bf \hat
  n}({\bf r})$ on the flow since it does not only change but it also
gives a contribution to the dissipative losses in the systems. In
particular, regions of high gradients of the director field result in
higher resistance to the flow. Such regions are mainly found around
disclinations, which are often unavoidable due to the geometry of the
system. If we consider, for example, perpendicular boundary conditions
on the surface of the sphere and a uniform director field far away
from it, there is a disclination loop around the sphere (see
fig.\ref{dir_field_force}), which is unavoidable for topological
reasons: The surface of the sphere corresponds to a $s=1$ point defect
and since the overall defect of the system must be zero (the director
field is uniform far from the particle) this defect must be balanced
by the disclination loop. The energy of the ring is roughly
proportional to its length, therefore it is favourable to have it as
small as possible. On the other hand, the rigid boundary conditions at
the surface of the particle push the ring away from the sphere, so
that the final position is given by the balance of the two effects.

The drag force is sensitive to the radius of this loop. In both
limiting cases (flow and director parallel/perpendicular to each
other) the resistance is increased, but the magnitude of the influence
is quite different. For the director perpendicular to the velocity the
flow is parallel to the ring and it acts like a plate moved in the
liquid crystal. For velocity and (general) director parallel to each
other the flow is perpendicular to the ring, which does not only
increase the cross-section of high director gradients around the ring
and the liquid crystal flow but it also has a further effect: a
certain amount of the liquid crystal has to flow through the gap
between ring and sphere, where the director lies in the plane of the
ring and is therefore locally perpendicular to the direction of the
flow. As a consequence, the anisotropy of the system, i.e. the ratio
of the drag forces, decreases with an increasing radius of the
disclination loop. That means that stronger boundary conditions lower
the anisotropy.

The theoretical problem of a liquid crystal flowing around a body has
been addressed before. Diogo \cite{diogo} assumed the velocity field
around the sphere to be the same as for an isotropic fluid and
calculated the drag force for different angles between the director
and the velocity.  Roman and Terentjev \cite{roman} obtained an
analytic solution for the flow velocity for a fixed uniform director
field, by an expansion in the anisotropy of the viscosity.  Recently a
group around Kneppe and Schneider gave solutions for the velocity of
the liquid crystal, assuming a uniform director field, independent of
the flow\cite{sphere}.

All these solutions have their deficiencies. None of them, for
instance, considered the distribution of the director field due to the
boundary conditions on the particle.  This will be done in this
article, where the results are also compared with various
approximations for the director ${\bf \hat n}({\bf r})$.

The article is organised as follows: After a brief introduction to the
basic equations of the hydrodynamics of liquid crystals which are
needed in the next section, we give a short description of the
numerical method we used to solve the equations of motion. Section IV
gives the director fields we used and explains the limits when they
are valid. The results for the drag properties and a comparison with
experimental data are given in section V and, finally, we conclude
with a discussion of possible experiments are given in chapter six.

\section{Basic Concepts}

In this section we give a brief summary of the nematic hydrodynamics
that are used in this work. For derivations of these equations we refer the
reader to the basic textbook \cite{deGennes}, see
also \cite{our_paper}. The stress tensor of a nematic liquid crystal
consists of three contributions. They are the hydrodynamic pressure
$p$, the viscous stress given by the tensor

\begin{equation}
 \sigma^\prime_{ij}=\alpha_1n_in_jn_kn_lA_{kl}+\alpha_2n_jN_i+
\alpha_3n_i
N_j+\alpha_4A_{ij}+\alpha_5n_jn_kA_{ik}+\alpha_6n_in_kA_{jk}.
\label{sigmaP}
\end{equation}
(here and below in this article we use the tensor index notation, i.e.
an index appearing twice in a product means a summation over this
index, and the short hand notation for gradients $B_{,j}\equiv
\nabla_jB$). Here $\alpha_i$ are the viscosity (Leslie) coefficients,
${\bf A}$ represents the symmetric part of the fluid velocity
gradients [$A_{ij}=\frac{1}{2}(v_{i,j}+v_{j,i})$] and the vector
$N_i=\dot n_i+\frac{1}{2}\left[{\bf \hat n} \times \mathtt{curl}
  \mathbf{v} \right]_i$ is the change of the director with respect to
the background fluid.  Finally, there is a static (elastic)
contribution due to the curvature of the director field
\begin{equation}
\sigma^e_{ij}=-Kn_{k,j}n_{k,i}   \ , 
\end{equation} 
given in the one constant approximation (Frank elastic constants
$K_1=K_2=K_3\equiv K$).

The director field is determined by the balance between the static
molecular field ${\bf h^o}=K\nabla^2{\bf \hat n}$ and the viscous
molecular field
$h^\prime_i=(\alpha_2-\alpha_3)N_i+(\alpha_6-\alpha_5)n_jA_{ij}$. The
total molecular field has to be parallel to the director but
$h^\prime$ can be neglected in the low Ericksen number
regime\cite{our_paper} ($Er=\alpha v R/K\ll1$, where $v$ is a
characteristic velocity and $R$ the radius of the sphere). This
condition is met in a typical thermotropic liquid crystal with $K \sim
10^{-11} \hbox{N}; \alpha \sim (5-10) \times 10^{-2} \hbox{Pa} \cdot
\hbox{s}$ in the case of $vR \ll 10^{-8}\hbox{m}^2 \, \hbox{s}^{-1}$
which allows speeds of millimetre per second for small colloid
particles ($R \sim 10 \mu m$).

Considering low Reynolds number flow and using the equation of continuity we end up with seven equations 

\begin{eqnarray}\label{motion}
\sigma_{ij,j}&=&0\label{motion1}\\
v_{i,i}&=&0\label{motion3}\\
K n_{i,jj}&=&\lambda n_i\label{motion2}
\end{eqnarray}
for seven unknown variables (three for the velocity field ${\bf v}$,
three for the director ${\bf \hat n}$ and the Lagrange multiplier
$\lambda$ constraining ${\bf \hat n}^2=1$, and one for the pressure).
The equation (\ref{motion2}) for the director is decoupled from
the velocity due to the low Ericksen number approach and can be solved
separately for the static problem, which then leaves only the
hydrodynamic part of Eqs.(\ref{motion1}-\ref{motion3}).

Once the velocity field $\mathbf{v}(\mathbf{r})$ and the pressure
$p(\mathbf{r})$ are obtained, the convenient way to determine the drag
force is by calculating the total dissipation in the system
\begin{equation}
  F\cdot v_\infty = \int \left(\mbox{\boldmath $\sigma^\prime$}:{\bf
      A}+{\bf h^\prime}\cdot {\bf N}\right) \quad dV
\end{equation}
where $v_\infty$ is the constant velocity of the fluid at infinity.

\section{Numerical Method}

We followed the example of Heuer, Kneppe and Schneider\cite{sphere}
and used the artificial compressibility method (see for example
\cite{peyret}) to solve the equations of motion. The idea of this
method is that the system starts with an arbitrary start-up velocity
and pressure field and relaxes in an artificial time towards its
equilibrium, which is the solution we are looking for
($\partial_tp=0$, $\partial_tv_i=0$).  The equations to solve are:

\begin{eqnarray}\label{eq_of_motion}
\sigma_{ij,j}&=&\partial_tv_i\nonumber\\
v_{i,i}&=&-c^2\partial_tp
\end{eqnarray}
where $c$ is an arbitrary damping parameter, which should be chosen as
large as possible to speed up the calculation (however, if $c$ is too
large the numerical scheme becomes unstable).

Due to the linearity of Eqs.(\ref{eq_of_motion}) it is necessary to
solve them only for the two particular cases where flow and director
are parallel and perpendicular to each other [${\bf \hat
n}(\infty)\parallel{\bf v}(\infty)$ and ${\bf \hat n}(\infty)\perp{\bf
v}(\infty)$]. The advantage of these solutions is the simple
geometry. For an arbitrary angle between velocity and director they
are just added together, i.e. the friction drag can be calculated by
the resistance tensor
$M_{ij}=M_\perp\delta_{ij}+(M_\parallel-M_\perp)n_in_j$, which
determines the response of the drag force on the sphere to the flow
around it:

$$F_i=M_{ij}({\bf \hat n})v_{\infty j}=M_\perp v_{\infty i} +
(M_\parallel - M_\perp )({\bf v_\infty \cdot\hat n})n_i.$$

For an isotropic liquid the tensor is simply $M_{ij}=M\delta_{ij}$
where the constant $M$ is given by the Stokes friction $M=-6\pi R\eta$.
The ratio $M_\perp/M_\parallel$ is a measure for the anisotropy in the
system since this gives the lift effect in the drag force. In the
first case, assuming that the flow is along the z-axis and the
director is parallel to it, the system is symmetric with respect to
azimuthal rotations around the z-axis. When the velocity components
are transformed to cylindrical components ($v_x,v_y,v_z\rightarrow
v_\rho,v_\phi,v_z$) the azimuthal velocity $v_\phi$ is zero everywhere
in the system and the problem becomes two-dimensional. Furthermore it
is favourable to use a spherical coordinate system with an inverse
radius ($\xi=\frac{1}{r}=\frac{1}{\sqrt{x^2+y^2+z^2}}$;
$\theta=\arctan\frac{\sqrt{x^2+y^2}}{z}$; $\phi=\arctan\frac{y}{x}$).
This has two advantages: the outer boundary conditions [${\bf
  v}(\infty)$, $p(\infty)$] are included in the grid used for the
calculations and the mesh size of the grid is smaller near the surface
of the sphere, where most of the changes happen, and large far from
the particle, where the values stay almost constant. It is sufficient
to pursue the calculations in one quadrant only since the other three
are given by symmetry. It is also evident that the radial velocity
must be zero at both boundaries. The values for $v_z$ at
$\theta=\frac{\pi}{2}$ can be computed as the inner grid points
whereas the values at $\theta=0$ request special treatment since they
contain the term cosec $\theta$ and are therefore of the form
``$0/0$''. Since it was not possible to obtain them by an
interpolation, we simplified the equation by taking the limit for
$\theta=0$ [$v_z(\xi,0)=\lim_{\theta\rightarrow0}v_z(\xi,\theta)$]
analytically (application of L'Hopital's rule).

In the second case, the director perpendicular to the velocity, there
is no rotational symmetry and the calculations have to be done on a
three dimensional grid. It is again favourable to use spherical
coordinates with an inverse radius (see fig.\ref{grid_3d}) for the
reasons explained above, but this time the velocity components are
kept Cartesian [i.e.
$v_x(\xi,\phi,\theta),v_y(\xi,\phi,\theta),v_z(\xi,\phi,\theta)$].
Due to the symmetry it is sufficient to solve the equations in one
octant.  The conditions on the boundaries of this octant and the
needed values are calculated as is shown in table 1.

The constant number of grid points in the plane of the azimuthal angle
$\phi$ direction (independent of $\theta$) leads to a decrease in the
mesh-size in real space while approaching the pole and finally yields
a non-uniqueness for the pole ($\theta=0$) itself. A constant distance
in real space would request fewer points (factor $\sim 0.7$) but it
involves more calculations since the derivatives become more
difficult. Therefore we chose the grid shown in fig.\ref{grid_3d}. The
values at the z-axis, which are non-unique, were then calculated for
each $\phi$ and set to their average over $\phi$ ($v_z(\xi,\phi,0)=
<v_z(\xi,\phi,0)>_\phi$ for every $\xi$).

\section{The Director Field}

As described above, the director field can be taken as fixed during
the calculations in the low Ericksen number regime. In order to study
the influence of simplifying assumptions concerning the form of the
field we performed the calculations with different director fields ${\bf
\hat n}({\bf r})$. In the one constant approximation the director
field is described by the minimum of the Frank free energy $F_d$
\begin{equation}\label{lapl_dir}
F_d=\int\left(\nabla\cdot{\bf \hat n} \right)^2+(\nabla{\bf \hat n})^2 dV\,\,.
\end{equation}
If we take into account that ${\bf \hat n}$ is a unit vector and set
${\bf \hat n}(\infty)$ parallel to the z-axis, we can write the
director components as
\begin{eqnarray}
n_x&=&\sin\beta\sin\gamma\\
n_y&=&\sin\beta\cos\gamma\\
n_z&=&\cos\beta\qquad\qquad
\end{eqnarray}
where $\beta$ and $\gamma$ are angles dependent on the spatial
coordinates. The director field in our problem is rotational symmetric
with respect to the z-axis. We can therefore set $\gamma=\arctan y/x$.
Inserting this in Eq.(\ref{lapl_dir}) and minimising it, we are left with
one equation for the polar angle $\beta({\bf r})$:
\begin{equation}\label{dir_eq}
\nabla^2\beta-\frac{\sin2\beta}{2r^2\sin^2\theta}=0
\end{equation}
where $\beta$ is a function of the radius $r$ and the azimuthal angle
$\theta$. There are several possibilities to proceed with finding the
static director field (see fig.3):

\begin{itemize}
\item If we neglect the boundary conditions on the surface (anchoring
  energy is zero) we get the (trivial) solution
  $\beta(r,\theta)\equiv 0$, i.e. the director is uniform in space,
  parallel to the to the z-axis. This was the approach chosen by the
  authors \cite{sphere} in their analysis.
\item Provided the anchoring on the surface is weak and therefore the
  angle $\beta$ [the deviation from ${\bf \hat
    n}(\infty)$] remains small, Eq.(\ref{dir_eq}) can be linearised and yields
\begin{equation}\label{ukr1}
\nabla^2\beta-\frac{\beta}{r^2\sin^2\theta}=0.
\end{equation}
This equation, with the corresponding boundary conditions and the symmetry of
the problem, can be easily solved and gives $\beta=\left(\frac{R
    W}{4 K}\right)R^3\sin 2\theta/r^3$, where $W$ is the anchoring
energy and $R$ the radius of the particle\cite{shiyanovskii}.

\item If we assume strong anchoring on the surface of the sphere
  ($WR/K\gg 1$) the director field is forced to have a disclination
  loop (radius $a$) around the equator of the field. The direction of
  the molecules in the plane of the ring must be radial between ring
  and disclination, and parallel to the z-axis in this plane outside
  the disclination. Furthermore the perturbation of the field must
  decay as $1/r^3$ far from the sphere. The simplest function which
  shows this behaviour is
\begin{equation}\label{ansatz}
\beta=\theta-\frac{1}{2}\arctan\frac{\sin 2\theta}{\cos 2\theta +\left(\frac{a}{r}\right)^3}.
\end{equation}

There is an extensive discussion of the features and details of the
director field in the strong anchoring regime \cite{shiyanovskii}. The
conclusion reached \cite{shiyanovskii} is that Eq.(\ref{ansatz})
provides a very good approximation, well describing the far-field
behaviour, the disclination ring vicinity, and even the weak
anchoring case when the ring radius $a$ is taken $a\rightarrow
WR^4/4K$.


\item There is no analytical solution for the whole problem
  (satisfying Eq.(\ref{dir_eq}) {\it and} the boundary conditions). We
  therefore solved the equilibrium equation (\ref{dir_eq}) numerically
  with a method similar to the artificial compressibility method
  mentioned before. In this way we obtain the exact director field
  ${\bf \hat n}({\bf r})$ on every point of our grid.
\end{itemize}

\section{results}
First we examine the influence of the different director fields
[uniform, trial function $\beta$ and the exact numerical ${\bf \hat
  n}({\bf r})$] on the drag force acting on the sphere, using the
particular set of viscous coefficients of MBBA\cite{deGennes}. As
expected, the uniform field shows a much lower drag force for both
principal configurations of ${\bf \hat n}(\infty)$ and $v(\infty)$
than the trial field and the exact field, see fig.4 (for the velocity
parallel to the director it is even smaller than for the isotropic
drag force).  On the other hand the anisotropy of the drag forces (the
ratio of the two forces $F_\perp$/$F_\parallel$) becomes smaller for
the realistic non-uniform field. This is due the following: the flow
velocity around the sphere is the highest in the region of the equator
plane perpendicular to the line of motion. If the liquid crystal is
oriented along the same z-axis we get a very high gradient of the
director in exactly the same region due to the effect of the
disclination ring. In the other principal configuration, where the
director ${\bf \hat n}(\infty)$ is perpendicular to the z-axis and,
therefore, perpendicular to the velocity, the disclination with its
high gradients is the same, but this time the loop is around a
longitude of the sphere. It still increases the drag but in a much
smaller region since the flow velocity at the stagnant poles is almost
zero already.  The anisotropy in the drag force for the three director
fields yield:
\begin{equation}
  \frac{F_\perp}{F_\parallel}|_{\sf
    uniform}=1.69\qquad\frac{F_\perp}{F_\parallel}|_{\sf
    trial \beta}=1.50\qquad\frac{F_\perp}{F_\parallel}|_{\sf
    exact}=1.50
\end{equation}

The results obtained with the trial field are surprisingly close to
those of the exact director field. They obviously reflect the
important features of the field which are mainly the disclination loop
and the 1/$r^3$ decay of the deviation in the angle far from the
particle, whereas the particular details near the particle and the
disclination seem to be of minor importance. Therefore, the drag force
is determined by the long-range effects. The difference between the
drag force of the trial functions and the drag force obtained from the
exact numerical solution is less than 1\% and the difference in the
force ratios is smaller than the accuracy of the calculations. In most
practical cases it should be sufficient accuracy to use these trial
${\bf \hat n}({\bf r})$ fields instead of a numerical solution of the
governing equations. The uniform field, on the other hand, is not a
very useful assumption since its resulting drag force differs from
that for the exact ${\bf \hat n}({\bf r})$ distribution by up to 20\%.
The significantly higher ratio is a particular problem since it shows
that the effect of boundary conditions on the surface cannot be simply
modelled by a larger effective hydrodynamic radius.  The dependence on
the mutual orientation of $n_\infty$ and $v_\infty$ is too strong.

These results can be compared with experimental figures. The diffusion
of particles in a nematic liquid crystals \cite{frenkel} is described
by the diffusion tensor ${\bf D}$ which is indirectly proportional to
the resistance tensor ${\bf D}=k_B T\left({\bf M}\right)^{-1}$, where
$k_B$ is the Boltzmann factor and T the temperature. Consequently, the
tensor is of the same form as the mobility tensor
$D_{ij}=D_\perp\delta_{ij}+(D_\parallel-D_\perp)n_in_j$ with $D_\perp=
k_B T/M_\perp$ and $D_\parallel=k_B T/M_\parallel$. Recent experiments
\cite{ohta} showed for the the self diffusion constants of MBBA a
ratio of $D_\parallel/D_\perp\sim1.5$. Easier to determine
experimentally is the anisotropy in the electric conductivity
\cite{heppke} $\mbox{\boldmath $\mu$}$ of a sample, which is related
to the diffusion by
\begin{equation}
{\bf D}=\frac{k_B T}{ne^2}\mbox{\boldmath $\mu$}
\end{equation}
for $n$ charge carriers of charge $e$ per cm$^3$. The conductance
anisotropy was often measured \cite{deGennes} and for MBBA it is usually
equal to $\mu_\parallel/\mu_\perp\sim1.5$. Both experiments are in
excellent agreement with our result for $M_\perp/M_\parallel=1.50$.

The dependence of the drag force on the temperature is also of great
interest in many experiments. The viscous coefficients
$\alpha_1,\alpha_2,\alpha_3,\alpha_5$ and $\alpha_6$, in the first
approximation, depend linearly on the order parameter S in the region
close to the nematic$\leftrightarrow$isotropic transition temperature
$T_{ni}$. The order parameter S itself can be approximated by Haller's equation \cite{haller}
\begin{equation}
S=\left(1-\frac{T}{T_{ni}}\right)^\gamma \,
\end{equation}
where $\gamma$ is determined experimentally for MBBA \cite{reifenrath} to be $\gamma=0.188$. 
The viscous coefficients scale, therefore:
\begin{equation}
\alpha_1\rightarrow \alpha_1*S;\quad\alpha_2\rightarrow \alpha_2*S;\quad\alpha_3\rightarrow \alpha_3*S;\quad\alpha_4\rightarrow \alpha_4;\quad\alpha_5\rightarrow \alpha_5*S;\quad\alpha_6\rightarrow \alpha_6*S.
\end{equation}

The drag force shows for the perpendicular case more or less the same
behaviour for all director fields: after a jump at the transition
point, it increases while lowering the temperature, in the beginning
rapidly, then slower and slower. On the other hand there is a
qualitative change for the parallel drag force: while it jumps to a
lower value and then decreases further for the uniform field, it shows
a small change to a higher value at the transition point at which it
stays almost constant independent of the temperature, both for the trial
field of $\beta$ and the exact field (see fig.4).

The boundary conditions are not absolutely rigid in many experiments
due to a finite anchoring energy, in which case the approximative
director field (\ref{ukr1}) can be used. A typical colloidal particle
of radius $R=10^{-5}$m in a liquid crystal with an elastic constant of
$K\sim10^{-11}\mbox{N}$ has an anchoring energy of $W\sim10^{-5}\dots
10^{-7}$J. That corresponds to a relevant dimensionless factor of
$WR/K=0.1\dots 10$. Our calculations showed a slow linear increase of
the drag drag forces $F_\perp$ and $F_\parallel$ in the range of weak
anchoring, $WR/K=0\dots 4$ due to the deviation of the director field
from the uniform state. This results in gradients in the field which
increase the dissipation and, therefore, the resistance of the
particle to the flow. The effect on the force in the parallel case
$F_\parallel$ is obviously stronger than in the perpendicular case
$F_\perp$ which is reflected in a linear decrease of the anisotropy
ratio $F_\perp/F_\parallel$ while enlarging the anchoring energy $W$.

The authors of \cite{shiyanovskii} have also examined the case of
charged particles, when the radial electric field near the surface
forces the disclination loop to be pushed further away from the particle.
An approximate expression for the loop radius $a$ is then given by
\begin{equation}
a^2=\frac{\epsilon_a q^2}{32\epsilon^2(5K+K_{13}-2 K_{24})}
\end{equation}
where $\epsilon_a$ is the dielectric constant, $q$ the charge of the
sphere, and $K_{13}$ and $K_{24}$ elastic constants (assuming no
immediate screening). One expects an increase in the drag forces and a
decrease in their ratio since the hydrodynamical effective cross
section of large director gradients is far more increased for the
parallel case than for the perpendicular one. For instance, taking the
loop radius $a\sim2R$ the results for the MBBA-set of Leslie
coefficients are:

\begin{equation}
  F_\perp=1.75\,F_{iso}\qquad
  F_\parallel=1.23\,F_{iso}\qquad\frac{F_\perp}{F_\parallel}=1.43
\end{equation}
where the drag forces are given in units of MBBA in the isotropic
phase $F_{iso}=-6\pi R\eta v$ using $\eta =0.5\alpha_4$ as viscosity
coefficient.  As mentioned above, the anisotropy of the drag forces
decreases with increasing the strength of the boundary conditions from
$F_\perp/F_\parallel$=1.69 for the uniform field (anchoring energy
$W=0$), followed by a slow linear increase for weak anchoring
($WR/K\ll 1$), and $F_\perp/F_\parallel$=1.5 for rigid anchoring to
$F_\perp/F_\parallel$=1.43 for the case of the charged particle, which
can be considered as ``over-strong'' anchoring.

The molecular characteristics of the liquid crystal are inherent in
the viscous coefficients. These coefficients depend, among other things, on the
shape of the molecules which form the liquid crystal. This influence
can be modelled by an affine transformation model \cite{ehrentraut}
giving the viscous coefficients depending on the molecular aspect
ratio $\ell_{\|}/\ell_{\bot}$:
\begin{eqnarray}
\alpha_1 &=& -\frac{1}{2} \alpha_0 \left(
\frac{\ell_{\|}}{\ell_{\bot}} -  \frac{\ell_{\bot}}{\ell_{\|}}
\right)^2 \ ; \qquad 
\alpha_2 = \frac{1}{2} \alpha_0 \left(1 - \left[
\frac{\ell_{\|}}{\ell_{\bot}} \right]^2 \right) \ ;  
\label{leslies} \\
\alpha_3 &=&  \frac{1}{2} \alpha_0 \left( \left[
\frac{\ell_{\bot}}{\ell_{\|}} \right]^2 -1 \right) \ ; \qquad
\alpha_4 = \alpha_4 \ ; \qquad 
\alpha_5 = -\alpha_2 \ ; \qquad \alpha_6 = \alpha_3 \nonumber
\end{eqnarray}

The authors of \cite{ehrentraut} determined the aspect ratio of MBBA
to be $\ell_{\|}/\ell_{\bot}=5/2$. We used this value to obtain the
constant $\alpha_0$ by comparing the largest coefficient $\alpha_2$
with the experimental value and the isotropic coefficient $\alpha_4$
was taken from MBBA directly. We calculated the cases of two
particular configurations: a rod-like molecule with ratio
$\ell_\parallel/\ell_\perp=7/2$ and a disc-like system
$\ell_\parallel/\ell_\perp=3/5$. The results of these calculations
are:
\begin{eqnarray}
 \mbox{rod-like molecule:}\qquad F_\perp=2.33\,F_{iso}\qquad
  F_\parallel=1.78\,F_{iso}\qquad\frac{F_\perp}{F_\parallel}=1.31\nonumber\\
 \mbox{disc-like molecule:}\qquad F_\perp=0.94\,F_{iso}\qquad
  F_\parallel=1.52\,F_{iso}\qquad\frac{F_\perp}{F_\parallel}=0.62
\end{eqnarray}
Note the inverted ratio of the drag forces for the disc-like molecules.

\section{Conclusion}
Considering the low Ericksen number regime ($Er=\alpha v R/K\ll1$) the
director field can be taken as independent of the flow in the first
approximation. We, therefore, took several static director fields
(approximations for the field in the limit of strong and weak
anchoring and the solution of the governing equations). The equations
of motion were then solved numerically for the different fields ${\bf
  \hat n}({\bf r})$ using the viscous coefficients of MBBA. Due to the
linearity of the Eqs.(\ref{eq_of_motion}) it is only necessary to solve
two limiting case, for director and velocity parallel and
perpendicular to each other at infinity [${\bf \hat n}(\infty)
\parallel {\bf v}(\infty)$ and ${\bf \hat n}(\infty) \perp {\bf
  v}(\infty)$]. This yields the drag forces $F_\parallel$ and
$F_\perp$ which can be combined for the general case by the mobility
tensor.

The comparison of the drag forces for the different director fields
showed that the disclination loop around the sphere, which is
topologically necessary for a large anchoring energy of the molecules
on the surface, does not only increase the forces itself but also
decreases their ratio to $F_\perp/F_\parallel=1.50$ compared to a
uniform case ($F_\perp/F_\parallel=1.69$). Trial director fields,
constructed from the basic features of the director field
(disclination ring and 1/$r^3$ decay of the far field), showed to be a
very good approximation. The difference between the values of
$F_\parallel$ and $F_\perp$ compared to the ones obtained for the
exact field is less than 1\%.

The temperature dependence of the drag force showed an increase in the
force $F_\perp$ for decreasing temperature and an almost constant
value for the parallel force ($F_\parallel$). An approximation for the
director field for weak anchoring energy shows a linear decrease for
lowering the anchoring energy in both particular drag forces as well
as in their their ratio.

The value for MBBA ($F_\perp/F_\parallel=1.50$), using the exact
solution of the director field, is in good agreement with experimental
results, measured by the static conductivity and the self diffusion of
MBBA (both $\sim1.5$).

The disclination loop can be pushed away from the sphere, for
instance, in the case of non-screened charges on the particle. This
increases the particular forces compared with the uncharged case,
where the loop is close to the surface, and yields for a loop radius
of twice the particle radius an even lower ratio of
$F_\perp/F_\parallel=1.43$.

The viscous coefficients of other materials can be approximated by an
affine transformation model, which uses the aspect ratio
($\ell_\parallel/\ell_\perp$) of the molecules as parameter. For a
rod-like molecule ($\ell_\parallel/\ell_\perp$=7/2) we obtained an
anisotropy in the drag force of $F_\perp/F_\parallel=1.31$ and for a
disc-like molecule ($\ell_\parallel/\ell_\perp$=3/5) the ratio
obtained was $F_\perp/F_\parallel=0.62$. The ratio smaller than one
indicates that the lift force turns the particle away from the
director whereas a ratio larger than one forces the particle in the
direction of the director.

It is especially interesting to examine the lift component of the drag
force, i.e. the non-dissipative force acting perpendicular to the line
of particle motion. It resembles magnetic forces and leads to physical
phenomena, similar to the Hall effect. In a long cell, the ratio
between the cross voltage $U^\ast$ and the applied voltage $U$ is
determined by the anisotropy ratio $V=\mu_\parallel/\mu_\perp$
\cite{heppke}
\begin{equation}
\frac{U^\ast}{U}=-\frac{b}{a}\frac{\sin 2\theta}{(V+1)/(V-1)-\cos 2\theta}
\end{equation}
where $a$ is the width of the sample (in direction of the applied
voltage) and $b$ the thickness of the sample (in direction of the
cross voltage). The conductivity is determined by the movement of the
charge carriers and, therefore, inversely proportional to the resistance
which yields for the anisotropy ratios 
$V=M_\perp/M_\parallel=F_\perp/F_\parallel$.

Since it is difficult to produce samples of liquid crystals without
disclinations, which are sufficiently large to perform measurements of
moving particles, the effect of the lift force (the component of the
drag force which is perpendicular to the driving force) could be
observed in a long cylinder. If the boundaries force the liquid
crystal to be perpendicular to the walls of the cylinder, the director
field will ``escape to the third dimension'', i.e. it will turn round
to be parallel to the long axis of the cylinder while approaching its
centre since this is energetically much more favourable than a
disclination line. If the distances with the same, well defined
curvature along the z-axis are large enough, little spheres, which are
dropped in the sample, should show a certain measurable displacement
during their way down in a gravitational field.

Further possibilities are the usage of electric and magnetic
fields. Moving particles can be guided by changing the director
orientation in the sample to direct them to a certain destination in
the sample. This enables the guiding of uncharged and unpolarizable
particles with electric or magnetic fields

\begin{center}
ACKNOWLEDGEMENTS
\end{center}

The authors have benefited from discussions with C. Bergemann, F.M.
Leslie, J.R. Melrose and M.Warner. This research has been supported by
the EPSRC UK and Unilever PLC.

\scalegraphlabel{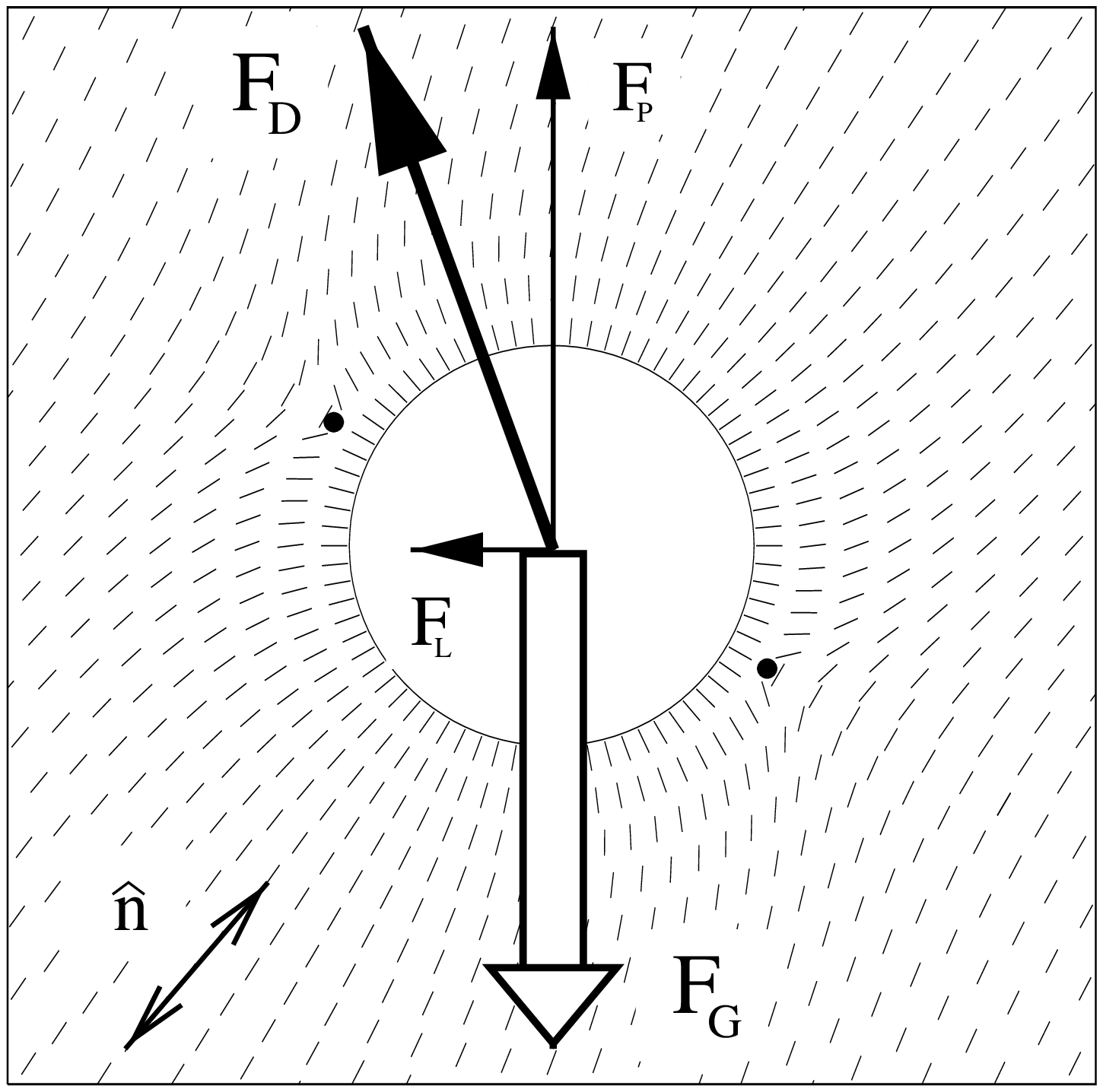}{18cm}{}{dir_field_force}

\newpage

\scalegraphlabel{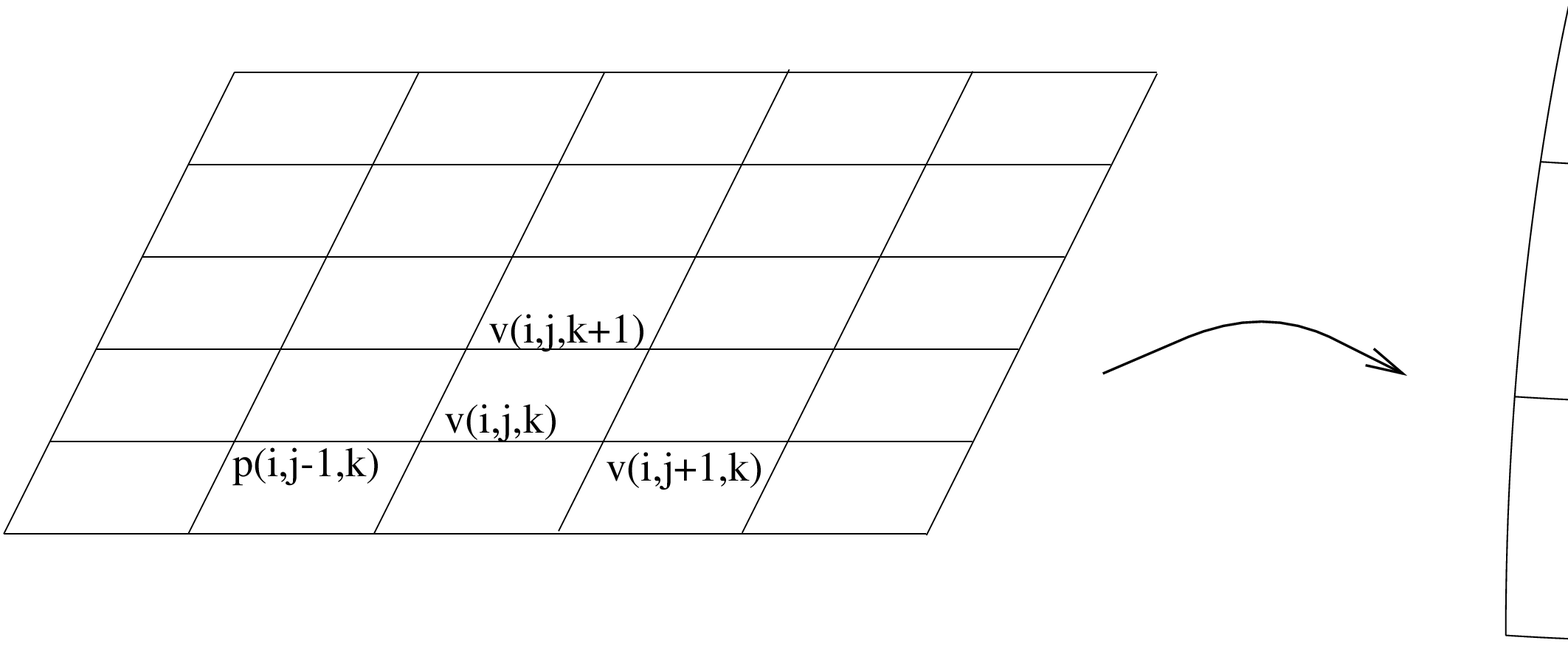}{5cm}{}{grid_3d}

\newpage

\scalegraphlabel{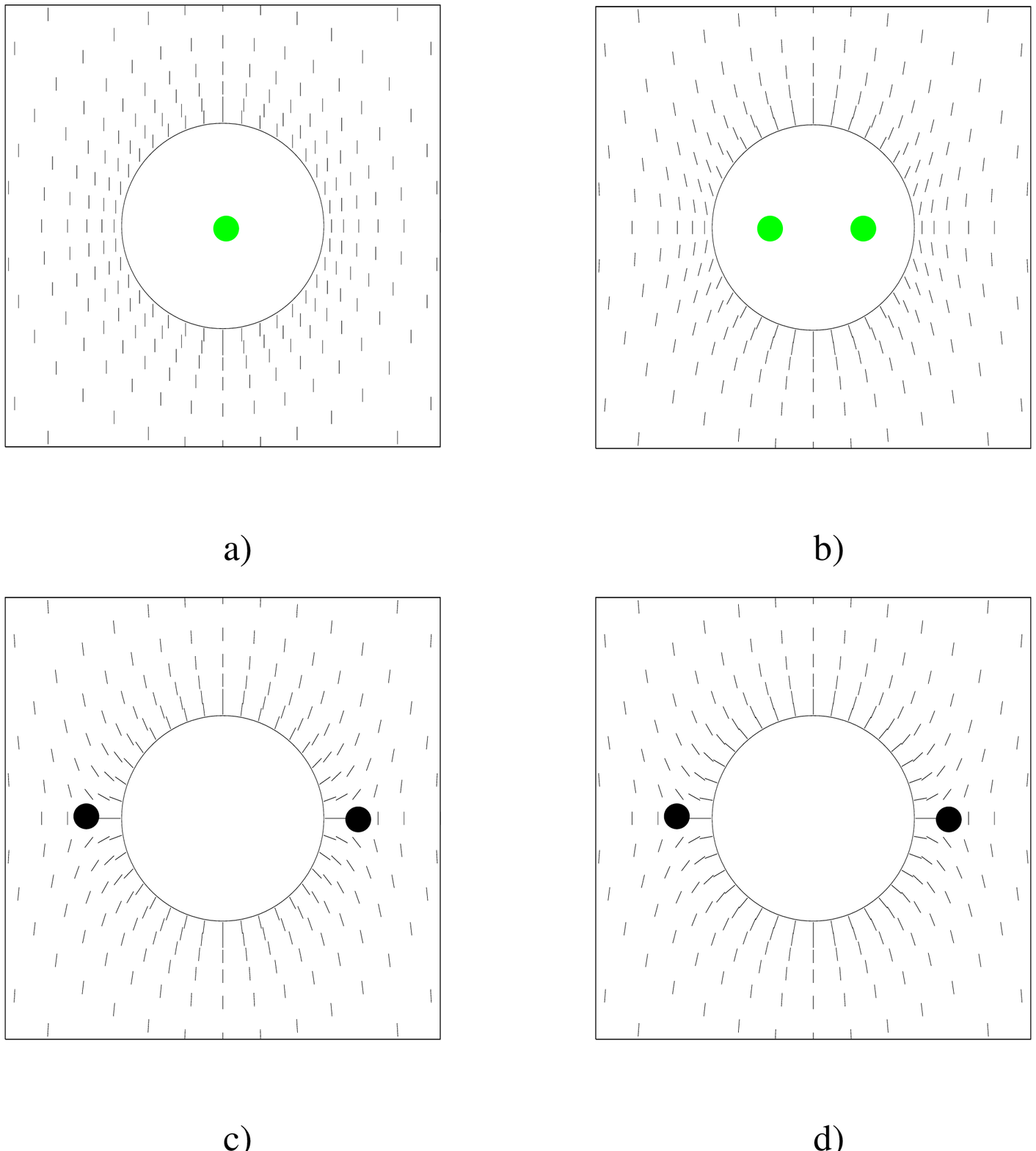}{16cm}
{}{dir_fields}

\newpage

\begin{figure}[htb]
\setlength{\unitlength}{0.1bp}
\special{!
/gnudict 40 dict def
gnudict begin
/Color false def
/Solid false def
/gnulinewidth 5.000 def
/vshift -33 def
/dl {10 mul} def
/hpt 31.5 def
/vpt 31.5 def
/M {moveto} bind def
/L {lineto} bind def
/R {rmoveto} bind def
/V {rlineto} bind def
/vpt2 vpt 2 mul def
/hpt2 hpt 2 mul def
/Lshow { currentpoint stroke M
  0 vshift R show } def
/Rshow { currentpoint stroke M
  dup stringwidth pop neg vshift R show } def
/Cshow { currentpoint stroke M
  dup stringwidth pop -2 div vshift R show } def
/DL { Color {setrgbcolor Solid {pop []} if 0 setdash }
 {pop pop pop Solid {pop []} if 0 setdash} ifelse } def
/BL { stroke gnulinewidth 2 mul setlinewidth } def
/AL { stroke gnulinewidth 2 div setlinewidth } def
/PL { stroke gnulinewidth setlinewidth } def
/LTb { BL [] 0 0 0 DL } def
/LTa { AL [1 dl 2 dl] 0 setdash 0 0 0 setrgbcolor } def
/LT0 { PL [] 0 1 0 DL } def
/LT1 { PL [4 dl 2 dl] 0 0 1 DL } def
/LT2 { PL [2 dl 3 dl] 1 0 0 DL } def
/LT3 { PL [1 dl 1.5 dl] 1 0 1 DL } def
/LT4 { PL [5 dl 2 dl 1 dl 2 dl] 0 1 1 DL } def
/LT5 { PL [4 dl 3 dl 1 dl 3 dl] 1 1 0 DL } def
/LT6 { PL [2 dl 2 dl 2 dl 4 dl] 0 0 0 DL } def
/LT7 { PL [2 dl 2 dl 2 dl 2 dl 2 dl 4 dl] 1 0.3 0 DL } def
/LT8 { PL [2 dl 2 dl 2 dl 2 dl 2 dl 2 dl 2 dl 4 dl] 0.5 0.5 0.5 DL } def
/P { stroke [] 0 setdash
  currentlinewidth 2 div sub M
  0 currentlinewidth V stroke } def
/D { stroke [] 0 setdash 2 copy vpt add M
  hpt neg vpt neg V hpt vpt neg V
  hpt vpt V hpt neg vpt V closepath stroke
  P } def
/A { stroke [] 0 setdash vpt sub M 0 vpt2 V
  currentpoint stroke M
  hpt neg vpt neg R hpt2 0 V stroke
  } def
/B { stroke [] 0 setdash 2 copy exch hpt sub exch vpt add M
  0 vpt2 neg V hpt2 0 V 0 vpt2 V
  hpt2 neg 0 V closepath stroke
  P } def
/C { stroke [] 0 setdash exch hpt sub exch vpt add M
  hpt2 vpt2 neg V currentpoint stroke M
  hpt2 neg 0 R hpt2 vpt2 V stroke } def
/T { stroke [] 0 setdash 2 copy vpt 1.12 mul add M
  hpt neg vpt -1.62 mul V
  hpt 2 mul 0 V
  hpt neg vpt 1.62 mul V closepath stroke
  P  } def
/S { 2 copy A C} def
end
}
\begin{picture}(3600,2160)(0,0)
\special{"
gnudict begin
gsave
50 50 translate
0.100 0.100 scale
0 setgray
/Helvetica findfont 100 scalefont setfont
newpath
-500.000000 -500.000000 translate
LTa
600 251 M
0 1858 V
LTb
600 251 M
63 0 V
2754 0 R
-63 0 V
600 457 M
63 0 V
2754 0 R
-63 0 V
600 664 M
63 0 V
2754 0 R
-63 0 V
600 870 M
63 0 V
2754 0 R
-63 0 V
600 1077 M
63 0 V
2754 0 R
-63 0 V
600 1283 M
63 0 V
2754 0 R
-63 0 V
600 1490 M
63 0 V
2754 0 R
-63 0 V
600 1696 M
63 0 V
2754 0 R
-63 0 V
600 1903 M
63 0 V
2754 0 R
-63 0 V
600 2109 M
63 0 V
2754 0 R
-63 0 V
600 251 M
0 63 V
0 1795 R
0 -63 V
952 251 M
0 63 V
0 1795 R
0 -63 V
1304 251 M
0 63 V
0 1795 R
0 -63 V
1656 251 M
0 63 V
0 1795 R
0 -63 V
2009 251 M
0 63 V
0 1795 R
0 -63 V
2361 251 M
0 63 V
0 1795 R
0 -63 V
2713 251 M
0 63 V
0 1795 R
0 -63 V
3065 251 M
0 63 V
0 1795 R
0 -63 V
3417 251 M
0 63 V
0 1795 R
0 -63 V
600 251 M
2817 0 V
0 1858 V
-2817 0 V
600 251 L
LT0
3114 1946 M
180 0 V
600 808 M
1434 35 V
120 0 V
101 -2 V
92 -22 V
600 1923 M
968 -169 V
176 -37 V
159 -42 V
131 -49 V
120 -70 V
53 -42 V
48 -51 V
43 -68 V
17 -37 V
8 -25 V
9 -33 V
15 -77 V
0 -559 R
894 0 V
LT1
3114 1846 M
180 0 V
600 796 M
1434 27 V
120 0 V
101 -1 V
92 -21 V
600 1892 M
968 -165 V
176 -35 V
159 -37 V
131 -52 V
120 -68 V
53 -41 V
48 -52 V
43 -64 V
17 -35 V
8 -25 V
9 -31 V
15 -76 V
LT2
3114 1746 M
180 0 V
600 354 M
2034 478 L
120 21 V
101 41 V
92 62 V
600 1531 M
2034 1325 L
120 -42 V
101 -62 V
43 -47 V
25 -41 V
24 -77 V
stroke
grestore
end
showpage
}
\put(3054,1746){\makebox(0,0)[r]{uniform field}}
\put(3054,1846){\makebox(0,0)[r]{trial function $\beta$}}
\put(3054,1946){\makebox(0,0)[r]{exact field}}
\put(2008,51){\makebox(0,0){effective temperature $T/T_{ni}$}}
\put(100,1180){%
\special{ps: gsave currentpoint currentpoint translate
270 rotate neg exch neg exch translate}%
\makebox(0,0)[b]{\shortstack{drag force (in units of $F_{iso}$)}}%
\special{ps: currentpoint grestore moveto}%
}
\put(3417,151){\makebox(0,0){1.6}}
\put(3065,151){\makebox(0,0){1.4}}
\put(2713,151){\makebox(0,0){1.2}}
\put(2361,151){\makebox(0,0){1}}
\put(2009,151){\makebox(0,0){0.8}}
\put(1656,151){\makebox(0,0){0.6}}
\put(1304,151){\makebox(0,0){0.4}}
\put(952,151){\makebox(0,0){0.2}}
\put(600,151){\makebox(0,0){0}}
\put(540,2109){\makebox(0,0)[r]{1.7}}
\put(540,1903){\makebox(0,0)[r]{1.6}}
\put(540,1696){\makebox(0,0)[r]{1.5}}
\put(540,1490){\makebox(0,0)[r]{1.4}}
\put(540,1283){\makebox(0,0)[r]{1.3}}
\put(540,1077){\makebox(0,0)[r]{1.2}}
\put(540,870){\makebox(0,0)[r]{1.1}}
\put(540,664){\makebox(0,0)[r]{1}}
\put(540,457){\makebox(0,0)[r]{0.9}}
\put(540,251){\makebox(0,0)[r]{0.8}}
\put(2008,-1051){\makebox(0,0){Fig.4 }}
\end{picture} 
\label{drag_force}
\end{figure}

\newpage 

\scalegraphlabel{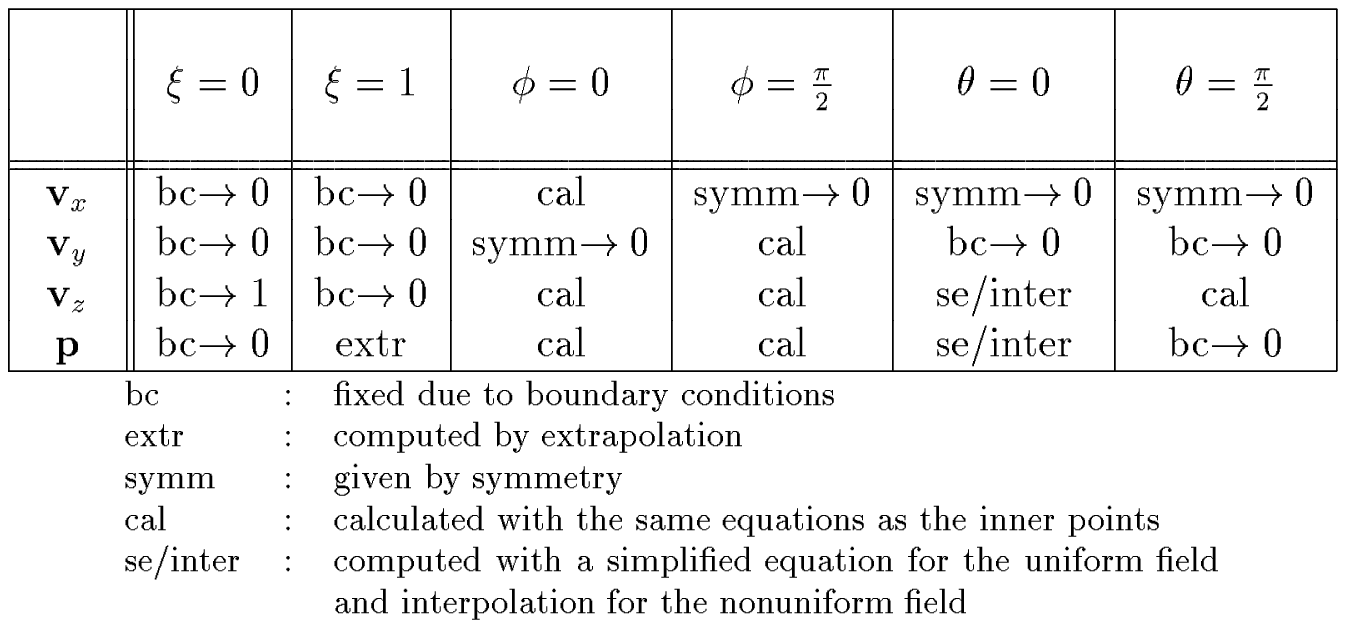}{16cm}{}{table1}

\newpage

Fig.1: If a sphere falls down in a gravitational field and the
director field is not parallel to the force, their is besides the
friction $F_p$ anti-parallel to the gravitational force also a
component $F_L$ perpendicular to it, the so-called lift force. Note
also the disclination loop around the particle due to the boundary
conditions, indicated by the dots.

\bigskip

Fig.2: One layer of the matrix ($\xi =\frac{1}{{\bf r}}=$const.) and
its transformation to real space. The lines of constant angle $\phi$
transform to longitudes and constant $\theta$ to latitudes. Note the
decreasing distance between two longitudes while approaching the pole.
This yields a non-unique point at $\theta=0$.

\bigskip

Fig.3: The
  director fields for a) no anchoring at the surface of the sphere b)
  weak anchoring c) arctan field d) numerical solution fulfilling the
  boundary conditions. The black dots show the position of the
  disclination loop, the grey dots show where they would be (having
  this director field without the sphere).

\bigskip

Fig.4: The drag force on the sphere for three different director
  fields (uniform, trial and real field) depending on the effective
  temperature T/T$_{ni}$. The upper lines are for the case of director
  and velocity parallel , the lower ones for director and velocity
  perpendicular to each other.

\bigskip

Table 1: The treatment of the boundaries around the first octant. The
notation bc refers to the fact that this value is fixed by the
boundary conditions of the problem, symm refers to given by symmetry,
the values given by cal can be calculated with the untreated equations
of motion, se/inter the value was determined by an analytically
simplified equation for the uniform field and by an interpolation for
the non-uniform field, finally extr means the value was calculated by
an extrapolation.

\end{document}